# SWE-NFI: Studying and Benchmarking Coding Agents for Non-Functional Improvements

Pengyu Xue
York University
Toronto, Ontario, Canada

He Yang Yuan
York University
Toronto, Ontario, Canada

Xin Wang
The Hong Kong University of Science and Technology (Guangzhou)
Guangzhou, Guangdong, China

Junkai Chen
Singapore Management University
Singapore, Singapore

Haonan Zhang
University of Waterloo
Waterloo, Ontario, Canada

Boyuan Chen
Independent Researcher
Toronto, Ontario, Canada

Zishuo Ding
The Hong Kong University of Science and Technology (Guangzhou)
Guangzhou, Guangdong, China

Zhenhao Li
York University
Toronto, Ontario, Canada

Weiyi Shang
University of Waterloo
Waterloo, Ontario, Canada

## Abstract

Although coding agents have achieved impressive performance on correctness-oriented benchmarks, their ability to make behavior-preserving non-functional improvements (NFIs) remains underexplored. In real-world software development, developers continuously improve software quality without changing observable behavior, yet existing benchmarks primarily evaluate functional correctness and provide limited support for assessing these non-functional improvements. In this paper, we present SWE-NFI, a benchmark for evaluating coding agents on NFIs beyond functional correctness. SWE-NFI contains 188 tasks constructed from real merged pull requests in open-source Python projects. We operationalize five developer-oriented NFI aspects into 92 executable rules and develop an evaluation suite that first applies task-specific functional preservation checks to assess whether key properties of the original code are preserved and then performs rule-based NFI evaluation. For outputs that pass these checks, we measure NFI improvement as the change in the normalized rule score for the task's target aspect relative to the original code. We evaluate eight commercial and open-source agent configurations and run each configuration five times on every task. At most 70.0% of valid agent outputs satisfy all task-specific functional preservation checks. Among outputs that pass these checks, NFI performance varies substantially across aspects. In particular, all evaluated configurations achieve mean Logic Patterns improvements between 0.0 and 1.3, below the Human Reference mean of 1.5. Our benchmark and findings provide a reproducible foundation for evaluating coding agents on non-functional improvements and identifying the NFI aspects where further advances are needed.

## 1 Introduction

Coding agents powered by large language models have become increasingly integrated into modern software engineering workflows. These agents are capable of assisting developers in a wide range of functionality-oriented software engineering tasks, such as resolving GitHub issues, fixing bugs, implementing feature requests [6, 8, 10, 29, 32, 63, 65, 74], and supporting emerging vibe coding workflows [9, 49].

In practice, real-world software development and maintenance involve far more than implementing new functionality or fixing bugs. Developers continuously make non-functional improvements (NFIs) to existing code, such as refining documentation [26], improving error handling [41], modernizing library usage [59], enhancing coding style [71], constructing system observability [16, 33, 60, 76], and simplifying program structure [46]. Although these changes typically preserve the original functionality, they play a critical role in improving software maintainability, readability, and long-term evolution [5, 40]. Consequently, NFIs constitute an indispensable part of software maintenance.

Despite the importance of NFIs, existing benchmarks for coding agents predominantly evaluate functionality-oriented tasks. Representative benchmarks typically evaluate the agents by comparing the generated code against expected behavior [4, 11, 29, 35]. Such evaluations have substantially advanced the development of coding agents [25, 34], but provide limited insights into their ability to perform NFIs. More importantly, functional correctness and NFIs capture complementary aspects of software quality. A functionally correct solution may still exhibit poor documentation, inadequate error handling, outdated library usage, or unnecessarily complex program structure. Since these issues affect maintainability and long-term evolution, evaluating coding agents solely by functional correctness gives an incomplete picture of practical usefulness.

To bridge this gap, we present SWE-NFI, a benchmark for systematically evaluating coding agents on NFIs beyond functional correctness. Following the five high-level aspects of non-functional improvements discussed in SWE-IF [75], we derive 92 executable rules from prior studies, real-world issues, and coding guidelines.

Based on these rules, we construct 188 benchmark tasks from real merged GitHub pull requests, including 145 single-file tasks and 43 multi-file tasks. For each task, an agent receives a natural-language description of the task and the original code and produces a revised version of the code. The corresponding real-world revision from the merged pull request serves as a reference for comparing during evaluation. Each task is accompanied by an evaluation suite consisting of task-specific functional preservation checks and rule-based NFI evaluation. The evaluation is fully automated using executable checks and does not rely on LLM-as-a-judge. Since the benchmark is constructed entirely from real merged pull requests, it captures the non-functional improvements that developers routinely perform during software development and maintenance.

We evaluate state-of-the-art commercial and open-source coding agents using SWE-NFI. Our results show that although the best-performing agent achieves a 70.0% functional preservation rate, all evaluated agents generally fall short of the corresponding human reference across multiple NFI aspects, especially for logic patterns, where the average improvement score ranges from 0.0 to 1.3 across all evaluated agents, compared with 1.5 for the human reference. We further find that multi-file tasks remain considerably more challenging than single-file tasks. In addition, we investigate the consistency of coding agents across repeated executions of the same task, as well as the practical costs of making non-functional improvements in terms of execution time and token consumption. The results suggest that, even when agent outputs pass the functional preservation checks, enabling agents to perform non-functional improvements remains challenging.

Our main contributions are summarized as follows:

- We operationalize NFIs into an executable evaluation framework that does not rely on LLM-as-a-judge. Specifically, following the five high-level NFI aspects proposed in prior work, we systematically derive 92 executable rules from multiple complementary sources, including prior studies, real-world open-source development practices, and coding guidelines.
- We construct SWE-NFI, a benchmark from real merged GitHub pull requests. The benchmark contains 188 NFI tasks, including 145 single-file and 43 multi-file. Each task is accompanied by an evaluation suite consisting of task-specific functional preservation checks and rule-based NFI evaluation.
- We conduct a comprehensive empirical study of state-of-the-art commercial and open-source coding agents on SWE-NFI. Our study provides an evaluation of coding agents on NFIs beyond functional correctness and discusses insights into their effectiveness, robustness, and practical deployment costs.

**Paper Organization.** This paper is organized as follows: Section 2 summarizes the related work, Section 3 describes the construction of our benchmark, Section 4 presents the experimental setup, Section 5 shows the experimental results, Section 6 discusses the case study and implications, Section 7 describes threats to validity, and Section 8 concludes the paper.

## 2 Related Work

### 2.1 Functional Code Generation Benchmarks

In the early days of large language models, code generation benchmarks, such as HumanEval [11] and MBPP [4], generally consist of method-level, self-contained programming problems with several test cases for functional evaluation. Motivated by these works, a series of follow-up benchmarks has emerged, improving upon them in various respects, such as task granularity (e.g., ClassEval [18], RepoEval [72]), language coverage (e.g., MultiPL-E [7], MBXP [3]), domain diversity (e.g., BigCodeBench [77], DS-1000 [30]), and test enhancement (e.g., EvalPlus [35]).

In particular, SWE-bench [29] advances the evaluation of code generation to a new stage: given an issue description, the model is required to edit multiple files to resolve it, with the resulting patch verified against comprehensive test suites. This work opens a new direction that more closely mirrors real-world development workflows, and has become the prevailing standard for evaluating the coding capabilities of LLMs and agents. It also inspires a number of works that extend and refine the original SWE-bench (e.g., feature addition [32], multilingual [70], and multimodal [66] extensions, among many others [14, 51, 69, 73]). Very recently, we have also observed that the generation of entire repositories or complete software systems [15, 24, 67] is nascent but developing rapidly. For example, ProgramBench [67] has attracted considerable attention from the community, which asks LLM agents to reproduce program executables and evaluate their function with agentic black-box fuzzing. Compared with these works focusing primarily on functional evaluation, SWE-NFI evaluates rule-measurable non-functional improvements to existing code.

### 2.2 Code Generation Evaluation Beyond Correctness

In addition to functional correctness, a growing number of studies examine other aspects of LLM-generated code. Among these, code efficiency has received considerable attention [17, 23, 27, 36, 37, 48, 50]. For example, EffiBench [27] collects efficiency-critical problems from LeetCode and compares the running time of LLM-generated solutions against human-written canonical ones. COFFE [48] uses computationally intensive test inputs and the number of CPU instructions for a more discriminative and precise evaluation of code efficiency. The security of code produced by LLMs and agents is another dimension beyond functionality [9, 45, 47, 53]. Pearce et al. [45] pioneer this direction by auditing GitHub Copilot completions and find that approximately 40% of them may be vulnerable. SecureVibeBench [9] reconstructs scenarios in which developers introduce security issues and uses them to evaluate agents. Beyond efficiency and security, more specialized aspects of code have been studied, such as portability [20, 59, 62], compliance [64], and fairness [19]. More closely related to our work, SWE-IF [75] augments programming problems from BigCodeBench [77] and LiveCodeBench [28] with non-functional instructions to evaluate code generation. In contrast, SWE-NFI is constructed from the modification histories of real software projects: each task originates from a merged pull request and pairs the original code with a real historical revision. This enables us to evaluate how coding agents improve existing project code in response to maintenance requests and to compare their measured improvements with the changes made by developers.

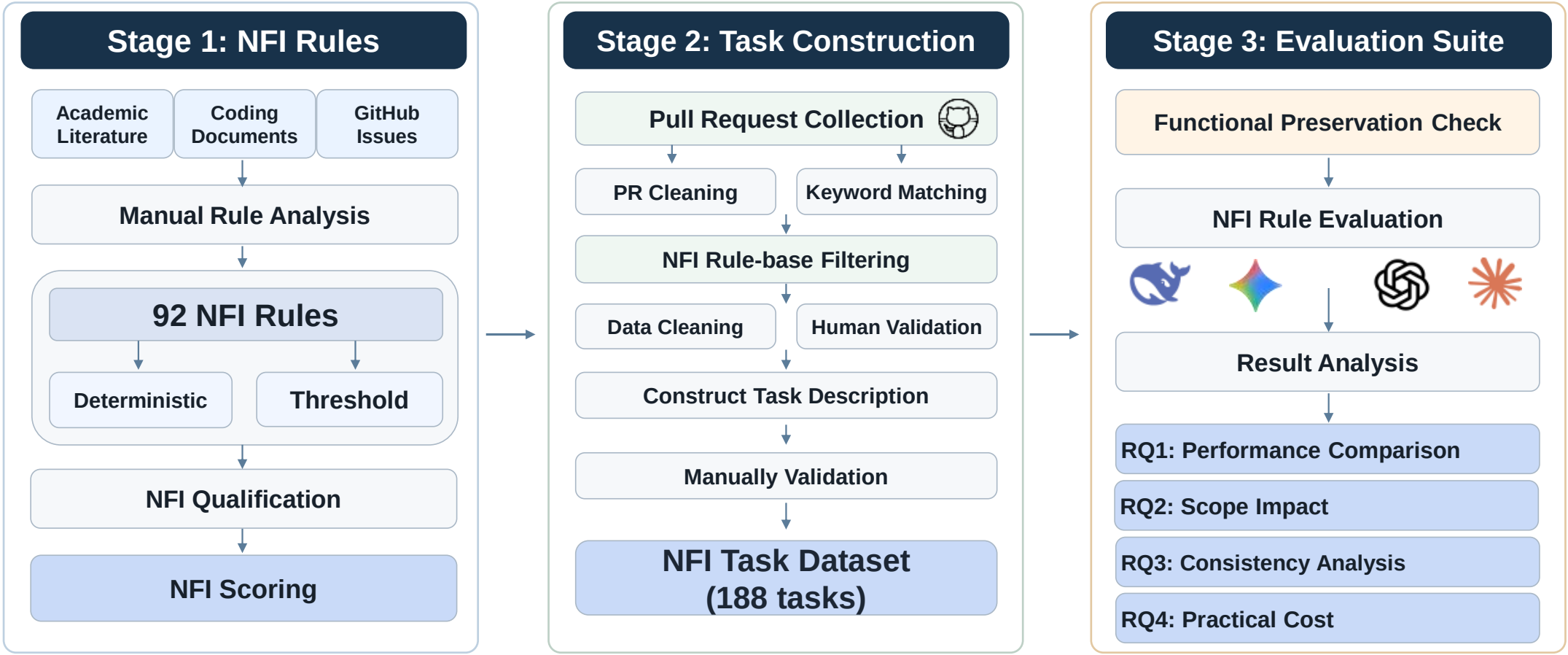

**Figure 1: Overview of SWE-NFI.**

## 2.3 LLM-Based Agents in Software Engineering

LLM-based agents have shown great capabilities in various software engineering tasks, including program repair, vulnerability discovery, and software refactoring [25, 34]. Bouzenia et al. [6] present the first autonomous LLM agent for program repair, namely RepairAgent. Built around tool use, this work designs a well-crafted tool set and employs dynamic templates together with a state machine to guide tool invocation for the LLM. AutoCodeRover [74] utilizes the reasoning ability of LLMs to retrieve code, localize relevant elements, and generate patches in an iterative manner. SWE-agent [65] proposes the idea of an Agent-Computer Interface, which transforms common computer operations into agent-friendly instructions to enable effective iteration between action and feedback. Building on SWE-agent, Abramovich et al. [1] design interactive agent tools and a context summarization strategy tailored for vulnerability discovery tasks. RefAgent [44] introduces a multi-agent framework for automatic code refactoring, where several specialized agents are responsible for roles such as planning and testing during the refactoring process.

Beyond academia, efforts from industry and the open-source community, exemplified by Claude Code [12], Codex [42], and OpenHands [61], have also joined in releasing and iterating on software engineering agents. As these agents become more capable, the need for a multi-dimensional evaluation beyond functionality alone becomes increasingly pressing, which is the target of SWE-NFI.

# 3 SWE-NFI Benchmark Framework

## 3.1 Benchmark Overview

SWE-NFI is designed to evaluate whether coding agents can make developer-oriented non-functional improvements beyond functional correctness. As shown in Figure 1, SWE-NFI consists of three main stages: **(1): NFI Rules:** We operationalize five aspects of NFIs into 92 executable rules; **(2) Task Construction:** We construct 188 benchmark tasks from real-world merged GitHub pull requests; **(3) Evaluation Suite:** We implement an evaluation suite that combines functional preservation checks with NFI rule-based evaluation. We illustrate each component in detail below.

## 3.2 NFI Rules

**Non-Functional Improvement Aspects.** Non-functional improvements cover a broad range of software quality concerns, such as maintainability, performance, security, reliability, and usability. In this work, we focus on developer-oriented, behavior-preserving improvements that are routinely performed during software maintenance. Specifically, following prior study [75], we study five selected aspects: ***Documentation***, ***Error Handling***, ***Library Constraints***, ***Coding Style***, and ***Logic Patterns***. Although these aspects provide a high-level characterization of non-functional improvements, they do not define executable evaluation criteria. Therefore, we further operationalize them into concrete rules that can be automatically evaluated. In this study, we focus on Python because it provides mature static-analysis tools and well-established coding standards, enabling reproducible benchmark construction and deterministic evaluation. For the five aspects, Documentation covers docstrings and type annotations; Error Handling addresses how code catches and propagates exceptions; Library Constraints pertains to API usage and deprecated interfaces; Coding Style dictates naming and formatting conventions; Logic Patterns evaluates control-flow complexity and coding patterns such as mutable default arguments.

**Rule Derivation.** We derive rules from three complementary sources: prior studies, open-source software artifacts, and Python coding guidelines.

First, we collect studies related to software maintainability and non-functional improvements from major software engineering venues (e.g., ICSE, FSE, ASE) and AI venues (e.g., AAAI, NeurIPS, ACL), published between 2020 and 2025. We retrieve candidate papers using aspect-specific keywords associated with the five NFI aspects, (eg., "static analysis", "linter", "code quality", "refactoring",

"code smells") resulting in 95 papers for Coding Style, Logic Patterns with 101 papers, Documentation with 78 papers, Error Handling with 97 papers, and Library Constraints with 39 papers. Following an in-depth manual screening of the literature, we narrowed the selection to 35 highly relevant papers that provided the theoretical foundation for our rule derivation.

Second, we collect real-world developer discussions from issues in widely used open-source Python projects. We start from the Top-100 starred Python projects in GitHub-Ranking and remove repositories that correspond to tutorials, demos, educational materials, instructions, or curated lists. We then use the GitHub Issue Search API to run aspect-specific keyword queries over the remaining repositories, restricting the search to issues created between 2020 and 2025 and removing duplicate URLs within each aspect. To focus on discussions with clear developer engagement, we keep only issues with at least 10 reactions and comments in total. This procedure yields 326 issue-aspect records from 314 distinct issue URLs across five NFI aspects: Coding Style (54), Logic Patterns (21), Documentation (104), Error Handling (31), and Library Constraints (116). We use these issues, together with the literature and coding guidelines, to derive candidate NFI rules.

Third, we incorporate coding guidelines that define recommended Python development practices. We systematically categorize and synthesize these guidelines into rules corresponding to our five evaluation aspects. Our rule set integrates standards from PEP 8 [58], PEP 3134 [68], PEP 585 [31], PEP 257 [21], and PEP 484 [57]. We also incorporate best practices from reputable technical sources, including the Tufts University coding standards [56], the pyupgrade documentation [54], and industry-standard metrics [52]. Each deterministic and threshold-based rule is derived from these official documents and specialized technical resources to ensure that our benchmark accurately reflects established professional coding standards.

**Manual Analysis and Refinement.** We then manually analyze the collected artifacts to identify concrete code improvement rules that can be mapped to the five NFI aspects. During the analysis, we focus on rules that satisfy two requirements. First, they should represent practices that frequently appear in prior studies or real-world software improvements. Second, they should be measurable through static analysis without requiring evaluation-time human annotation or LLM-based judgment.

The first author initially extracted candidate rules from the collected data. The extracted rules were then reviewed and refined by a second author. Disagreements were resolved through discussion until a consensus was reached. The Cohen's $\kappa$ [39] value in this process is 0.73, which indicates substantial agreement.

**Rules Summary.** Through this process, we derive 92 executable rules across the five NFI aspects. These rules transform high-level NFI aspects into concrete and measurable evaluation targets. To provide a clear overview of our evaluation mechanism, Table 1 reports the number of rules in each aspect, and Table 2 summarizes representative rules for each NFI aspect, categorized by their type of measurement and requirement. Due to space constraints, we list a portion of example rules in the table.

**NFI Rules Overview.** Through this process, we derive 92 executable rules across the five NFI aspects. We further organize these rules into two categories according to their measurement characteristics: *Deterministic Rules* and *Threshold-based Rules*:

- **Deterministic rules** correspond to rules that can be directly verified using static analysis without additional calibration. These rules are further divided into *Required Rules* and *Prohibited Rules*. Required rules describe coding practices that should be present in the generated code (e.g., function names should follow *snake_case*), whereas prohibited rules describe undesirable patterns that should be avoided (e.g., wildcard imports should not be used). Each deterministic rule is converted into a binary indicator (i.e., 1 if the code complies with the rule and 0 otherwise).
- **Threshold-based Rules** quantify code quality using numerical thresholds. Unlike deterministic rules, these rules require comparing a measured value against a predefined threshold. Examples include maximum function length, maximum nesting depth, cyclomatic complexity, and the number of parameters per function. In our benchmark, the cyclomatic complexity threshold of 10 follows McCabe's recommended upper limit [38], while thresholds without an established numerical standard are operational choices made for these evaluation rules.

**Table 1: Classification of the 92 released executable rules for Non-Functional Improvements evaluation.**

| Aspect | Deterministic | | Threshold-based | Total |
|---|---|---|---|---|
| | Required | Prohibited | Quantitative | |
| Documentation (Doc) | 9 | 5 | 5 | 19 |
| Error Handling (EH) | 3 | 7 | 2 | 12 |
| Library Constraints (LC) | 1 | 20 | 1 | 22 |
| Coding Style (CS) | 4 | 14 | 7 | 25 |
| Logic Patterns (LP) | 0 | 8 | 6 | 14 |
| **Total** | **17** | **54** | **21** | **92** |

**Non-Functional Improvements Quantification.** For each NFI aspect $p$, let $P_p(c)$ denote the number of rules satisfied by code version $c$ under aspect $p$, and let $T_p$ denote the total number of rules for aspect $p$. We first define the normalized rule score of code version $c$ as:

$$S_p(c) = \frac{P_p(c)}{T_p}. \quad (1)$$

Therefore, $S_p(c) \in [0, 1]$.

Given two code versions $c_{\text{before}}$ and $c_{\text{after}}$, we define the normalized NFI improvement under aspect $p$ as:

$$\begin{aligned} I_p(c_{\text{before}}, c_{\text{after}}) &= S_p(c_{\text{after}}) - S_p(c_{\text{before}}) \\ &= \frac{P_p(c_{\text{after}}) - P_p(c_{\text{before}})}{T_p}. \end{aligned} \quad (2)$$

A positive value indicates an improvement under NFI aspect $p$, zero indicates no measurable change, and a negative value indicates a regression.

For benchmark instance $i$, let $c^i_{\text{before}}$ denote the original code and let $c^i_{\text{after}}(x)$ denote the completed code produced by source $x$. For dataset instances, $x = h$ denotes the human reference implementation; during agent evaluation, $x = a$ denotes the output generated

**Table 2: Rules catalogue for the five NFI aspects. A dash indicates that no executable rules are defined.**

| Aspect | Measurement | Requirement | Example / Threshold | Explanation |
|---|---|---|---|---|
| **Documentation** | Deterministic | Required | Modules, classes, and methods | Docstring presence enforcement |
| | Deterministic | Prohibited | No empty docstrings, no TODOs | Content quality verification |
| | Threshold-based | Quantitative | Docstring/Type hint presence rate ≥ 80% | Documentation completeness targets |
| **Error Handling** | Deterministic | Required | Mandatory 'raise X from e' | Exception chaining enforcement |
| | Deterministic | Prohibited | No bare 'except:', no silent drops | Safety and exception robustness |
| | Threshold-based | Quantitative | Try-block LOC ratio ≤ 0.5, handlers ≤ 3 | Exception scope management |
| **Library Constraints** | Deterministic | Required | Use pathlib, collections.abc | Modern API migration |
| | Deterministic | Prohibited | No tf.Session, np.bool, .ix[] | Legacy API prevention |
| | Threshold-based | Quantitative | Deprecated API count ≤ 2 per file | API usage frequency monitoring |
| **Coding Style** | Deterministic | Required | snake_case, PascalCase, Line break after operators, UPPER_CASE | Naming conventions enforcement |
| | Deterministic | Prohibited | No trailing whitespace, mixed tabs | Formatting standards maintenance |
| | Threshold-based | Quantitative | Line length ≤ 100, magic numbers ≤ 5 | Complexity & style density control |
| **Logic Patterns** | Deterministic | Required | — | — |
| | Deterministic | Prohibited | No mutable default args, no 'while True' | Anti-pattern avoidance |
| | Threshold-based | Quantitative | Cyclomatic complexity ≤ 10, depth ≤ 3 | Complexity and logic density |

by agent $a$. We compute the NFI improvement of source $x$ under aspect $p$ as:

$$I_{i,p}(x) = I_p\left(c^i_{\text{before}},\, c^i_{\text{after}}(x)\right). \tag{3}$$

We report all NFI improvement values after scaling $I_{i,p}(x)$ to a percentage scale. Here, $I_{i,p}(h)$ denotes the human reference improvement, and $I_{i,p}(a)$ denotes the improvement achieved by agent $a$.

The scaled NFI improvement expresses the percentage-point change in the proportion of satisfied rules under aspect $p$. It measures improvement relative to the original code, while $S_p(c^i_{\text{after}}(x))$ denotes the normalized rule score of the revised code for instance $i$ and source $x$.

## 3.3 Benchmark Dataset Construction

We construct the SWE-NFI dataset through a multi-stage process. We collect merged pull requests from widely used open-source Python repositories and use the aspect-specific rule checkers introduced in Section 3.2 to identify code changes with measurable NFI improvements. We then apply automated filters and manually validate the remaining changes to exclude cases that are ambiguous or unsuitable as self-contained tasks. For each selected change, we write a task description based on the developer's intent without revealing the final code or the evaluation rules. The resulting benchmark contains 188 tasks: 145 single-file tasks and 43 multi-file tasks.

**Step 1: Real-world Pull Request Collection.** We construct SWE-NFI from real-world merged pull requests in widely used open-source Python repositories. We first collect repositories using the GitHub Search API and filter them by programming language, repository status, and popularity. We exclude forks, archived repositories, tutorials, demos, educational projects, and curated lists, yielding 100 production-grade Python repositories, each with at least 45,000 stars (median 66,254). We then collect merged pull requests from 2020 to 2025. To focus on developer-driven non-functional improvements, we remove bot-generated pull requests, dependency updates, release pull requests, version bumps, and revert commits. We further apply aspect-specific keyword matching to increase the likelihood of collecting modifications related to the five NFI aspects. The basic unit of data collection is a PR-file pair, where each record corresponds to one merged pull request and one modified Python file. After deduplication, we obtain 226,955 raw records.

**Step 2: NFI Rule-Based Filtering.** We next identify records with measurable non-functional improvements by computing the NFI improvement score using Eq. (3). We retain only records satisfying $I_{i,p}(h) > 0$. This process yields 6,775 positive-improvement records, 211,914 records without measurable changes, 7,566 regressions, and 700 scorer errors. Only the positive-improvement records are retained for subsequent benchmark construction. The merged `code_after` provides evidence that the task admits a measurable improvement under our rule-based evaluation, but it does not represent an optimal solution or a human-performance ceiling. During evaluation, each agent receives the task description and `code_before`, and its revision is scored independently using the same evaluation rules.

**Step 3: Data Cleaning and Human Validation.** We further clean and validate the remaining records to construct benchmark task candidates. We first apply automatic filters, including repository blocklists, diff size and file length constraints, per-repository caps, and SHA-256 deduplication, resulting in 1,215 candidate records for manual review.

Two authors independently review each candidate to verify that it is focused and represents a genuine non-functional improvement. We further exclude instances requiring substantial external context, large-scale migrations, public API changes, or semantic changes. Disagreements are resolved through discussion until consensus is reached ($\kappa = 0.71$). This process yields 484 validated instances for subsequent benchmark construction.

**Step 4: Task Construction.** For each validated instance, we manually construct a natural-language task description that converts the original pull request into an NFI request. To preserve benchmark integrity, the description omits the final aspect, executable rules, code patches, and the reference implementation while preserving the original development intent. One author drafts each task description based on the pull request context and code changes, and a second author verifies and refines it for clarity, and self-containment. Disagreements are resolved through discussion until consensus is reached. The benchmark contains both single-file and multi-file tasks. Single-file tasks modify one file, whereas multi-file tasks require coordinated modifications across related files sharing a common improvement objective. After a final selection that discards non-self-contained or statically uncheckable instances, we obtain 188 tasks from 48 repositories.

## 3.4 Evaluation Suite

The evaluation suite of SWE-NFI consists of two complementary components: functional preservation checks and NFI rule-score evaluation. We first apply task-specific functional preservation checks to the generated source code. NFI evaluation is performed only on valid outputs that pass all corresponding checks. The entire process is fully automated and does not rely on LLM-as-a-judge or human annotation.

**Functional Preservation Checks.** For each benchmark task, we implement functional preservation checks to assess whether the revised code preserves key functional properties of the original code after performing the NFI task. One author drafts the checks, and a second author verifies and refines them for relevance to the task. Disagreements are resolved through discussion until consensus is reached. We validate the checks against the human reference implementations obtained from the corresponding merged pull requests; all 188 references pass their respective checks. A revision that passes all checks satisfies the requirements encoded by the evaluation suite and is therefore eligible for NFI scoring.

**NFI Rule-Score Evaluation.** We implement rule-based checkers using Ruff, Radon, AST-based analyses, and custom scripts to evaluate the 92 executable rules introduced in Section 3.2. Ruff detects coding-style violations and library usage issues, Radon measures structural properties (e.g., cyclomatic complexity and function length), AST analyses identify Python-specific patterns, and custom scripts implement rules unsupported by existing tools.

> **Benchmark Summary:** SWE-NFI operationalizes five NFI aspects into 92 executable rules and contains 188 real-world tasks, including 145 single-file and 43 multi-file tasks. Each task is accompanied by an evaluation suite combining functional preservation checks and rule-based NFI evaluation.

# 4 Experimental Setup

## 4.1 Studied Agents

We evaluate eight coding agent configurations, including two commercial coding agents and two open-source coding agents paired with three different foundation models.

**Commercial Coding Agents.** We study OpenAI Codex [42] and Claude Code Agent [12], two commercial coding agents designed for software engineering tasks. In our experiments, Codex is configured with *gpt-5.3-codex*, while Claude Code Agent uses *claude-sonnet-4-6*.

**Open-source Coding Agents.** We follow prior work [9] and study two open-source coding agents, Aider [2] and mini-SWE-agent [55], the officially recommended successor to SWE-agent [65]. To examine the influence of different foundation models, we equip both agents with the same three backbone models: *gpt-4.1* [43], *gemini-2.5-flash* [22], and *deepseek-chat* [13].

## 4.2 Experimental Protocol

To reduce the impact of stochasticity, we set the temperature to zero whenever the corresponding agent supports temperature control. For each agent configuration and each benchmark task, we repeat the experiment five times and report the average results across all runs. All agents are given the same task input, consisting of the natural-language task description and the corresponding *code_before*. Agents are expected to generate a revised version of the code based on the task description. The *code_after* is hidden from the agent's input and is used only as a reference for evaluation.

## 4.3 Results Analysis Protocol

**Output Validity.** Before evaluating the code generated by agents, we first verify whether the generated output is valid. An output is considered valid if it is non-empty and syntactically parseable Python. We report output validity first because an agent that fails to produce valid code cannot be meaningfully evaluated by our evaluation suites.

We report *valid-output rate*, which measures the fraction of agent runs where the generated output is both non-empty and syntactically valid Python. Agent runs with empty output or syntactically invalid Python output are excluded from subsequent functional preservation checks and NFI rule-score analysis.

**Functional Preservation Pass (FPC) Rate.** We apply the task-specific functional preservation checks to each valid output. An output proceeds to NFI scoring only if it satisfies all requirements specified by these checks.

**NFI Improvement.** For each valid agent run that passes the functional preservation checks, we compute the NFI improvement score under the target aspect $p$ using Eq. (3). In this evaluation setting, the source $x$ in Eq. (3) is instantiated as the evaluated agent $a$, and $c^i_{\text{after}}(a)$ denotes the evaluated output generated by that agent. The human reference improvement $I_{i,p}(h)$ is pre-computed from the dataset by instantiating $x = h$, where $c^i_{\text{after}}(h)$ denotes the human reference implementation. Both quantities are computed relative to the same *code_before*, and the human reference implementation is never exposed to agents.

**Aggregate Analysis.** We aggregate NFI results over valid agent runs that pass the functional preservation checks. Unless otherwise specified, reported NFI values are average improvements computed from $I_{i,p}(a)$ in Eq. (3) and scaled to a percentage. For aspect-level analysis, we compute averages separately for each NFI aspect. For subset-level analysis, such as single-file versus multi-file tasks, we compute averages over the corresponding evaluated outputs. If an agent has no valid outputs passing the functional preservation checks for a given subset, its NFI improvement for that subset is removed from calculation rather than treated as zero.

# 5 Results

In this section, we systematically evaluate coding agents on SWE-NFI and analyze their ability to perform tasks on non-functional improvements. Specifically, we investigate the following research questions:

- **RQ1:** How well do coding agents perform on non-functional improvements compared with human reference improvements?
- **RQ2:** How does task scope affect the ability of coding agents to make non-functional improvements?
- **RQ3:** How consistent are coding agents in producing non-functional improvements across repeated runs?
- **RQ4:** What is the practical cost of making non-functional improvements in terms of execution time and token consumption?

## RQ1: How well do coding agents perform on non-functional improvements compared with human reference improvements?

**Motivation.** Modern coding agents have demonstrated strong capabilities in achieving functional correctness. However, it remains unclear whether these agents can also produce measurable non-functional improvements. In professional software development, these aspects are critical for long-term maintainability and usability. RQ1 evaluates whether agent-generated revisions produce valid outputs that pass functional preservation checks and achieve measurable NFI. We compare these aspect-level improvements with human-reference improvements from real pull requests.

**Approach.** For RQ1, we evaluate each agent on the 188 tasks in SWE-NFI. We first report valid output rate and functional preservation pass rate to distinguish syntactically valid outputs from revisions satisfying the functional preservation checks. For NFI performance, each aspect column in Table 3 reports the average NFI improvement defined in Eq. (3). For an agent $a$, the reported value is the average of $I_{i,p}(a)$ over evaluated outputs under aspect $p$. For the Human row, the reported value is the average of $I_{i,p}(h)$, where *code_after* denotes the human reference implementation.

**Results.** Table 3 shows the results of this RQ. For each aspect, a "+" or "-" following the number indicates whether an agent's average improvement is above or below the Human Reference mean for that aspect. These comparisons are descriptive.

**Agent performance varies substantially across NFI aspects, with Logic Patterns presenting a common challenge.** Table 3 shows that the highest functional preservation pass rate is 70.0%, while NFI improvements vary considerably across the five aspects. All evaluated configurations report mean Logic Patterns improvements below the Human Reference mean, ranging from 0.0 to 1.3 compared with 1.5 for the Human Reference. In contrast, several configurations achieve higher mean improvements in Documentation, Error Handling, Library Constraints, or Coding Style.

**Several agent configurations achieve relatively strong improvements on specific NFI aspects.** Codex achieves a valid-output rate of 99.3% and reports mean improvements above the Human Reference means for Coding Style, Error Handling, and Library Constraints. Claude Code Agent achieves the highest functional preservation pass rate of 70.0% and reports higher mean improvements for Error Handling and Library Constraints. Aider + GPT-4.1 achieves the highest Documentation improvement among the evaluated configurations at 8.1, compared with the Human Reference mean of 8.0, and also reports a higher mean improvement for Error Handling. These results indicate that relative performance depends strongly on the target NFI aspect rather than on a single configuration consistently performing best across all aspects.

**Functional preservation is a major bottleneck for the Mini-SWE configurations.** Although the three Mini-SWE configurations produce syntactically valid outputs at relatively high rates of 83.8% to 98.4%, only 4.8% to 11.5% of their valid outputs pass the functional preservation checks. Consequently, substantially fewer Mini-SWE outputs are eligible for subsequent NFI evaluation. This contrast shows that producing valid code alone does not imply that the generated revision preserves the task-specific properties required for NFI evaluation.

> **RQ1 Summary:** Agent performance varies substantially across NFI aspects, and several configurations achieve higher mean improvements than the Human Reference on individual aspects. However, all evaluated configurations report lower mean Logic Patterns improvements than the Human Reference (0.0 - 1.3 vs. 1.5), highlighting logic-pattern improvement as a shared challenge.

## RQ2: How does task scope affect the ability of coding agents to make non-functional improvements?

**Motivation.** Real-world non-functional improvement tasks vary in scope. Some require localized modifications within a single file, whereas others involve coordinated changes across multiple files. Since modern coding agents are increasingly expected to handle repository-level development tasks, it is important to understand whether task scope influences their ability to make non-functional improvements. RQ2 compares agent performance on single-file and multi-file tasks to investigate how task scope affects NFI improvements across different quality aspects.

**Approach.** We split SWE-NFI into its single-file and multi-file subsets. For each agent and each subset, we compute the average NFI improvement using Eq. (3) with $x = a$. The reported values are computed over valid evaluated outputs passing the functional preservation checks only. For multi-file tasks, rule scores are aggregated across the modified files before computing task-level NFI improvement. Table 4 reports aspect-level improvements and their average across aspects. The $S$–$M$ columns subtract the multi-file average from the single-file average. Negative values indicate higher average improvement on multi-file tasks, while positive values indicate higher average improvement on single-file tasks.

**Results. The effect of task scope differs substantially across NFI aspects.** Table 4 shows that moving from single-file to multi-file tasks does not uniformly reduce NFI improvement. Documentation improvement increases for several configurations, including Claude Code Agent from 6.9 to 9.3 and Aider + Gemini 2.5 from 6.4 to 9.9. This pattern indicates that some documentation-oriented improvements can extend across multiple related files without reducing the measured NFI gain.

**Table 3: Performance comparison of coding agents on non-functional improvements and functional preservation checks (RQ1).**

| Agent | Valid (%) | FPC Rate (%) | Doc | EH | LC | CS | LP |
|---|---|---|---|---|---|---|---|
| Codex | 99.3 | 64.8 | $6.8^{-}$ | $2.1^{+}$ | $1.6^{+}$ | $2.0^{+}$ | $1.3^{-}$ |
| Claude Code Agent | 93.7 | 70.0 | $7.4^{-}$ | $2.1^{+}$ | $1.4^{+}$ | $1.0^{-}$ | $0.7^{-}$ |
| Aider + GPT-4.1 | 96.4 | 66.3 | $8.1^{+}$ | $2.3^{+}$ | $-1.1^{-}$ | $0.1^{-}$ | $0.9^{-}$ |
| Aider + DeepSeek | 73.9 | 67.0 | $6.0^{-}$ | $2.5^{+}$ | $0.5^{-}$ | $1.2^{-}$ | $0.7^{-}$ |
| Aider + Gemini 2.5 | 95.8 | 59.1 | $7.0^{-}$ | $2.1^{+}$ | $-0.5^{-}$ | $0.6^{-}$ | $0.7^{-}$ |
| Mini-SWE + GPT-4.1 | 98.4 | 4.8 | 0.0 | 0.0 | 0.0 | 0.0 | 0.0 |
| Mini-SWE + DeepSeek | 83.8 | 11.5 | $2.7^{-}$ | $1.6^{-}$ | $2.1^{+}$ | $1.0^{-}$ | $0.6^{-}$ |
| Mini-SWE + Gemini 2.5 | 91.1 | 5.3 | 0.0 | 0.0 | $0.5^{-}$ | $0.9^{-}$ | 0.0 |
| **Human Reference** | 100.0 | 100.0 | **8.0** | **2.0** | **1.1** | **1.5** | **1.5** |

**Table 4: Average NFI improvement by task scope (RQ2). Single-file (S) vs. multi-file (M) tasks.**

| Agent | Improvement (Single-file %) | | | | | | Improvement (Multi-file %) | | | | | | S − M (Difference %) | | | | | |
|---|---|---|---|---|---|---|---|---|---|---|---|---|---|---|---|---|---|---|
| | Doc | EH | LC | CS | LP | Avg. | Doc | EH | LC | CS | LP | Avg. | Doc | EH | LC | CS | LP | Avg. |
| Codex | 6.8 | 2.3 | 1.2 | 2.0 | 1.5 | 2.8 | 6.8 | 1.1 | 3.3 | 1.9 | 0.4 | 2.7 | 0.0 | +1.2 | −2.1 | +0.1 | +1.1 | +0.1 |
| Claude Code Agent | 6.9 | 2.4 | 1.1 | 1.1 | 0.7 | 2.4 | 9.3 | 0.7 | 2.5 | 0.7 | 0.5 | 2.8 | −2.4 | +1.7 | −1.4 | +0.4 | +0.2 | −0.4 |
| Aider + GPT-4.1 | 7.9 | 2.6 | −1.1 | 0.4 | 0.9 | 2.1 | 9.2 | 1.1 | −1.3 | −1.0 | 0.6 | 1.7 | −1.3 | +1.5 | +0.2 | +1.4 | +0.3 | +0.4 |
| Aider + DeepSeek | 6.1 | 2.7 | 0.3 | 1.4 | 0.7 | 2.2 | 5.6 | 1.5 | 1.7 | 0.7 | 0.4 | 2.0 | +0.5 | +1.2 | −1.4 | +0.7 | +0.3 | +0.2 |
| Aider + Gemini 2.5 | 6.4 | 2.3 | −0.5 | 0.6 | 0.9 | 1.9 | 9.9 | 1.2 | 0.1 | 0.4 | −0.1 | 2.3 | −3.5 | +1.1 | −0.6 | +0.2 | +1.0 | −0.4 |
| Mini-SWE + GPT-4.1 | 0.0 | 0.0 | 0.0 | 0.0 | 0.0 | 0.0 | -- | -- | -- | -- | -- | -- | -- | -- | -- | -- | -- | -- |
| Mini-SWE + DeepSeek | 2.4 | 2.0 | 1.9 | 1.1 | 0.9 | 1.7 | 4.0 | 0.8 | 2.7 | 0.9 | 0.0 | 1.7 | −1.6 | +1.2 | −0.8 | +0.2 | +0.9 | 0.0 |
| Mini-SWE + Gemini 2.5 | 0.0 | 0.0 | 0.5 | 0.9 | 0.0 | 0.3 | -- | -- | -- | -- | -- | -- | -- | -- | -- | -- | -- | -- |

**Error Handling and Logic Patterns consistently become more difficult in multi-file tasks.** Across all configurations with evaluable outputs in both settings, mean improvements decrease for these two aspects when moving from single-file to multi-file tasks. For example, Codex's Logic Patterns improvement decreases from 1.5 to 0.4, while Claude Code Agent's Error Handling improvement decreases from 2.4 to 0.7. The decreases reach up to 1.7 percentage points for Error Handling and 1.1 percentage points for Logic Patterns.

**Some configurations fail to produce eligible outputs for multi-file tasks.** Mini-SWE + GPT-4.1 and Mini-SWE + Gemini 2.5 produce no valid multi-file outputs that pass the functional preservation checks and therefore have no multi-file NFI scores. This result indicates that task scope can affect whether an agent reaches the NFI evaluation stage in addition to affecting the magnitude of the measured improvement.

> **RQ2 Summary:** Task scope affects NFI aspects differently. Multi-file tasks remain more challenging than single-file tasks for current coding agents. While documentation occasionally benefits from additional cross-file context, multi-file tasks consistently reduce improvements in coding style (up to 1.4), error handling (up to 1.7), and logic patterns (up to 1.1).

## RQ3: How consistent are coding agents in making non-functional improvements across repeated executions of the same task?

**Motivation.** Coding agents driven by large language models may exhibit stochastic behavior across repeated executions, even under the same benchmark task and agent configuration. For software maintenance tasks, stable behavior matters at multiple stages: producing valid outputs, passing functional preservation checks, and achieving measurable non-functional improvement. RQ3 evaluates whether these stages remain stable across repeated runs.

**Approach.** For RQ3, we use five repeated executions of each agent configuration to measure run-to-run consistency. We analyze three metrics: valid-output rate, functional preservation pass rate, and aggregate agent NFI improvement. The aggregate improvement, denoted as $\overline{I}_{\mathrm{NFI}}(a)$, is computed using Eq. (3) with $x = a$, over valid evaluated outputs passing the functional preservation checks. For each agent and metric, we report the five run values, their average, and standard deviation. Lower standard deviation indicates more stable behavior across repeated runs on the same tasks.

**Results. NFI improvements are generally stable across repeated runs.** Table 5 shows that Codex, Claude Code Agent, and the three Aider configurations exhibit relatively small run-to-run variation in aggregate NFI improvement. Their standard deviations range from 0.1 to 0.4 across the five repeated executions. For example, Claude Code Agent achieves an average aggregate NFI improvement of 13.0 with a standard deviation of 0.2, while Aider + GPT-4.1 achieves 13.6 with a standard deviation of 0.3.

**Most run-to-run variation occurs before NFI evaluation.** Valid-output rates and functional preservation pass rates vary more noticeably for some configurations than their measured NFI improvements. Mini-SWE + DeepSeek, for example, has a standard deviation of 3.1 in valid-output rate and 1.4 in functional preservation pass rate, compared with 1.4 in aggregate NFI improvement. Aider + DeepSeek similarly varies more in output validity and functional preservation than in its NFI score. These results indicate that repeated executions more strongly affect whether outputs become eligible for NFI evaluation than the measured improvement of outputs that reach this stage.

**High consistency does not necessarily indicate effective NFI performance.** Mini-SWE + GPT-4.1 produces identical aggregate NFI scores of 0.0 across all five runs, while Mini-SWE +

**Table 5: Results over five repeated runs (RQ3).**

| Agent | Metric | Run 1 | Run 2 | Run 3 | Run 4 | Run 5 | Avg. | Std. |
|---|---|---|---|---|---|---|---|---|
| **Codex** | Valid (%) | 98.4 | 100.0 | 99.5 | 98.9 | 99.5 | 99.3 | 0.5 |
| | FPC Pass (%) | 62.7 | 68.6 | 63.6 | 64.5 | 64.7 | 64.8 | 2.0 |
| | $\bar{I}_{\mathrm{NFI}}(a)$ | 12.4 | 12.5 | 13.4 | 12.6 | 12.6 | 12.7 | 0.4 |
| **Claude Code Agent** | Valid (%) | 97.3 | 92.0 | 94.0 | 92.0 | 93.1 | 93.7 | 2.0 |
| | FPC Pass (%) | 67.1 | 71.0 | 69.8 | 70.1 | 71.9 | 70.0 | 1.6 |
| | $\bar{I}_{\mathrm{NFI}}(a)$ | 13.1 | 13.1 | 12.8 | 12.8 | 13.4 | 13.0 | 0.2 |
| **Aider+GPT** | Valid (%) | 96.3 | 96.8 | 95.7 | 96.8 | 96.3 | 96.4 | 0.4 |
| | FPC Pass (%) | 66.1 | 67.4 | 67.1 | 65.8 | 65.0 | 66.3 | 0.9 |
| | $\bar{I}_{\mathrm{NFI}}(a)$ | 13.7 | 13.7 | 13.3 | 13.3 | 14.0 | 13.6 | 0.3 |
| **Aider+DS** | Valid (%) | 76.0 | 74.0 | 72.9 | 75.6 | 71.3 | 73.9 | 1.8 |
| | FPC Pass (%) | 66.2 | 69.4 | 67.9 | 65.0 | 66.3 | 67.0 | 1.5 |
| | $\bar{I}_{\mathrm{NFI}}(a)$ | 11.6 | 11.7 | 12.0 | 11.5 | 11.8 | 11.7 | 0.1 |
| **Aider+Gemini** | Valid (%) | 96.8 | 96.3 | 95.3 | 94.7 | 95.8 | 95.8 | 0.7 |
| | FPC Pass (%) | 59.2 | 59.4 | 60.2 | 58.5 | 58.3 | 59.1 | 0.7 |
| | $\bar{I}_{\mathrm{NFI}}(a)$ | 11.7 | 12.4 | 11.9 | 12.0 | 12.9 | 12.2 | 0.4 |
| **Mini-SWE+GPT** | Valid (%) | 98.4 | 98.4 | 98.4 | 98.4 | 98.4 | 98.4 | 0.0 |
| | FPC Pass (%) | 4.8 | 4.8 | 4.8 | 4.8 | 4.8 | 4.8 | 0.0 |
| | $\bar{I}_{\mathrm{NFI}}(a)$ | 0.0 | 0.0 | 0.0 | 0.0 | 0.0 | 0.0 | 0.0 |
| **Mini-SWE+DS** | Valid (%) | 79.8 | 87.6 | 85.7 | 85.7 | 80.3 | 83.8 | 3.1 |
| | FPC Pass (%) | 10.7 | 12.0 | 11.1 | 13.8 | 9.8 | 11.5 | 1.4 |
| | $\bar{I}_{\mathrm{NFI}}(a)$ | 8.6 | 6.6 | 5.5 | 7.8 | 5.0 | 6.7 | 1.4 |
| **Mini-SWE+Gemini** | Valid (%) | 91.5 | 90.4 | 91.5 | 92.5 | 89.4 | 91.1 | 1.1 |
| | FPC Pass (%) | 5.1 | 5.1 | 5.6 | 4.9 | 5.7 | 5.3 | 0.3 |
| | $\bar{I}_{\mathrm{NFI}}(a)$ | 0.9 | 0.9 | 0.8 | 0.9 | 0.8 | 0.9 | 0.0 |

Gemini 2.5 varies only between 0.8 and 0.9. However, their average functional preservation pass rates are only 4.8% and 5.3%, respectively. Their low variance therefore reflects consistently limited performance rather than consistently effective NFI improvement.

> **RQ3 Summary:** Aggregate NFI improvements are generally stable once outputs pass the validity and functional preservation checks. Most observed run-to-run variation occurs before NFI evaluation, and low variance alone does not indicate effective non-functional improvement.

## RQ4: What is the practical cost of making non-functional improvements in terms of execution time and token consumption?

**Motivation.** Beyond NFI performance, the practical utility of coding agents also depends on their computational cost. An agent that achieves strong NFI improvements may require substantially more execution time or token consumption than alternative configurations. RQ4 evaluates the computational cost of different coding agents and examines the relationship between resource consumption and NFI performance.

**Approach.** For each agent configuration, we measure execution time and token consumption. Execution time covers the period from task initiation to completion, including model inference and agent execution overhead; token consumption is the total number of input and output tokens. In each benchmark repetition, we sum these costs across all 188 task executions and report the average totals over five repetitions. We also calculate time and tokens per successful task to account for the resources spent on unsuccessful attempts. Specifically, for each repetition, we divide the total cost of all 188 task executions by the number of tasks whose outputs are valid, pass the functional preservation checks, and achieve a positive NFI improvement. In the results, we refer to such tasks as successful tasks. We then average the resulting ratios over the five repetitions.

**Results. Execution cost varies by orders of magnitude across agent configurations.** Figure 2a shows substantial differences in both execution time and token consumption. Claude Code Agent has the highest measured cost, averaging 14,330 seconds and 50.7 million tokens per benchmark repetition, followed by Mini-SWE + DeepSeek with 11,780 seconds and 32.1 million tokens. In contrast, Codex and all three Aider configurations consume fewer than one million tokens per repetition on average. Mini-SWE + GPT-4.1 has the lowest measured cost, requiring 816 seconds and 327 thousand tokens per repetition.

**Higher token consumption does not consistently correspond to stronger NFI performance.** Claude Code Agent consumes approximately 100 times more tokens than Codex and more than 50 times more than Aider + GPT-4.1, while its functional preservation pass rate is only 5.2 and 3.7 percentage points higher, respectively. Its higher resource consumption also does not translate into uniformly higher aspect-level NFI scores. Aider + GPT-4.1 reports higher Documentation (8.1 vs. 7.4) and Error Handling improvements (2.3 vs. 2.1), while Codex reports higher Coding Style (2.0 vs. 1.0) and Library Constraints improvements (1.6 vs. 1.4) than Claude Code Agent.

**Cost per successful task reveals the impact of unsuccessful attempts.** Figure 2b accounts for the resources spent per successful task. Here, a successful task refers to task whose outputs are valid, pass the functional preservation checks, and achieve a positive NFI improvement. Including the costs of unsuccessful attempts, we find that Codex requires 21.08 seconds and 4,339 tokens per successful improvement, while Claude Code Agent requires 120.02 seconds and 424,970 tokens. Despite its low total cost, Mini-SWE + GPT-4.1 achieves no successful improvements. These results show that practical efficiency depends on how often resource consumption translates into a measurable NFI improvement, not just on the cost of running an agent.

> **RQ4 Summary:** Execution time and token consumption vary substantially across agent configurations, and higher resource consumption does not consistently yield stronger NFI improvements. Accounting for unsuccessful attempts reveals even larger differences in the cost per successful task.

# 6 Discussion

## 6.1 Case Study

Figure 3 shows a task from SWE-NFI that is related to improving the aspect of *Documentation*. The task requires adding type annotations and docstrings to improve maintainability and static-analysis support without changing runtime behavior. For readability, the figure shows abbreviated versions of the task description and source code.

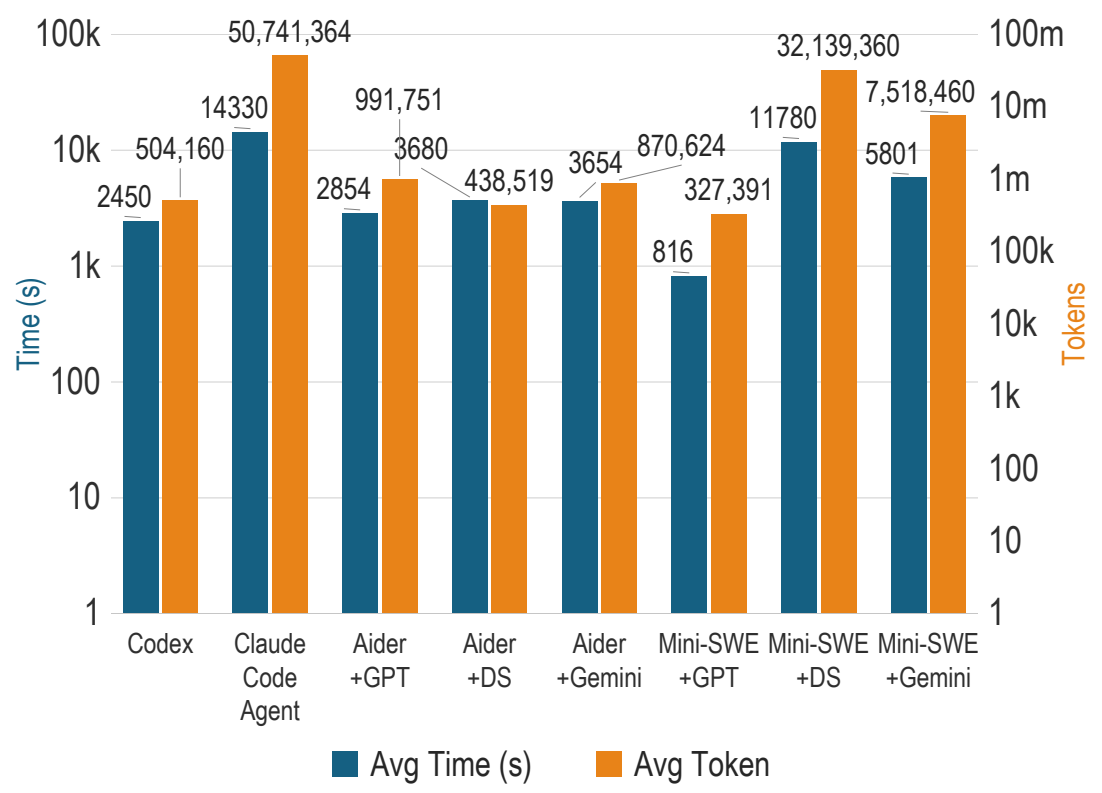

**(a) Total execution time and token usage.**

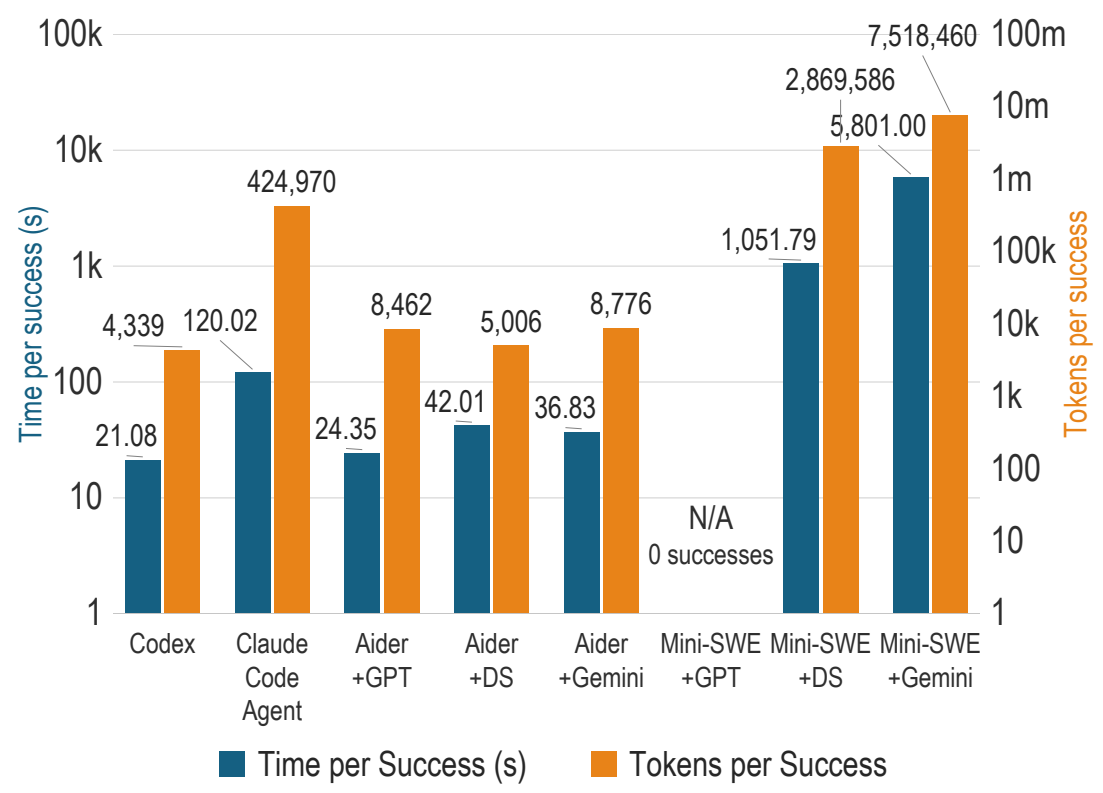

**(b) Cost per successful task.**

**Figure 2: Execution costs across agent configurations. (a) Total cost per benchmark repetition. (b) Total cost across all tasks divided by the number of successful tasks.**

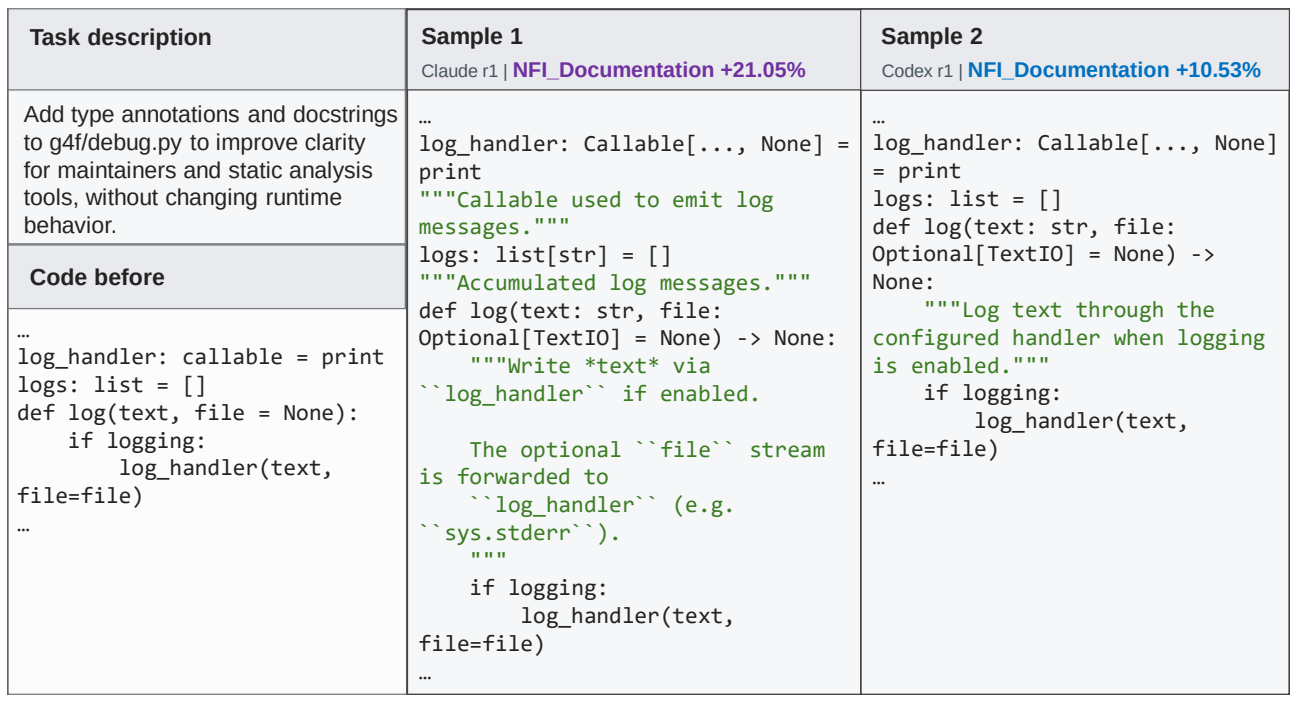

**Figure 3: Comparison of different Agent outputs**

Both agent outputs pass the task-specific functional preservation checks and achieve different levels of improvement in the Documentation rule score. Claude provides more comprehensive documentation, introduces more precise type annotations (e.g., *list[str]*), and documents the logging behavior in greater detail, resulting in a 21.05 *Documentation* rule-score improvement. Codex also adds type annotations and docstrings, but its revisions are less comprehensive, achieving a 10.53 *Documentation* improvement. This example illustrates that agents may produce different valid revisions with different priorities on the same task.

## 6.2 Implications

We discuss the implications of our findings for practitioners (P1-P3) and researchers (R1-R4), respectively.

**Implication P1: Use NFI rules to identify and review improvement opportunities.** The 92 executable rules translate recurring maintenance practices into checks that practitioners can apply to existing code. They can help identify opportunities to improve documentation, error handling, library usage, coding style, and logic patterns, and provide concrete criteria for reviewing proposed revisions. Because the rules capture selected, measurable practices, developers should interpret violations in the context of the project rather than treat every flagged instance as a change that must be made.

**Implication P2: NFI capability is aspect-dependent.** The aspect-level results show that coding agent performance varies substantially across NFI aspects. Documentation and error handling yield relatively stronger improvements for several configurations, whereas all evaluated agents report mean logic-pattern improvements below the Human Reference mean. Library-constraint performance also varies across configurations, including negative average improvements among outputs that pass the functional preservation checks. These differences suggest that progress on one type of non-functional improvement does not necessarily generalize to other aspects. Practitioners should therefore consider the target NFI aspect when selecting an agent and review its output for aspect-specific shortcomings.

**Implication P3: Consider execution cost when selecting agents for NFI tasks.** RQ4 reveals substantial differences in execution time and token consumption across agent configurations. Greater resource use does not consistently yield larger NFI improvements: on task runs where both agents pass the functional preservation checks and improve the target aspect, Claude Code Agent takes considerably longer than Codex while achieving a similar mean improvement. Practitioners should therefore weigh execution cost alongside functional preservation and NFI improvement when choosing an agent for non-functional maintenance.

**Implication R1: Current coding agents still facing challenges in non-functional improvements.** The benchmark instances in SWE-NFI are constructed from real merged pull requests and reflect non-functional improvements that developers make during software maintenance. These changes show that documentation, error handling, library usage, coding style, and logic patterns are recurring concerns in real-world development. However, several evaluated agent configurations achieve lower mean improvements than the Human Reference means across multiple NFI aspects. These results indicate that passing the functional preservation checks does not necessarily lead to consistent non-functional improvement across the measured aspects. Future research should therefore consider non-functional improvement as a distinct capability when developing and evaluating coding agents.

**Implication R2: Deterministic evaluation enables reproducible benchmarking.** Our benchmark operationalizes non-functional improvements using 92 executable rules. This enables deterministic, reproducible, and scalable evaluation without requiring evaluation-time human annotation. Although deterministic rule-based evaluation cannot capture every aspect of software quality, it provides a transparent foundation for comparing coding agents under the specified rules and is also flexible to extend.

**Implication R3: SWE-NFI can support the development of NFI-aware coding agents and tools.** The five-aspect framework and 92 executable rules provide a structured basis for evaluating the selected non-functional aspects of software. By converting maintenance practices into measurable criteria, SWE-NFI can support the development and analysis of coding agents and software engineering tools for documentation, error handling, library usage, coding style, and logic-pattern improvements. Beyond benchmark evaluation, these rules may also serve as explicit signals for designing agent feedback, validation, or iterative refinement mechanisms.

**Implication R4: Multi-file NFI tasks require better coordination across changes.** RQ2 shows that task scope affects NFI aspects differently. Among configurations with evaluable outputs in both subsets, measured improvements in Error Handling and Logic Patterns decrease from single-file to multi-file tasks, and some configurations produce no multi-file outputs eligible for NFI scoring. These findings motivate research on how agents identify related code locations and coordinate revisions across files while satisfying the functional preservation checks. Further evaluation on larger multi-file subsets is needed, particularly for underrepresented aspects.

## 7 Threats to Validity

**Construct Validity.** SWE-NFI operationalizes non-functional improvements using 92 executable NFI rules derived from multiple sources, including prior studies, open-source artifacts, and coding guidelines. Although these rules cannot capture all dimensions of software quality, we intentionally focus on improvements that can be measured automatically and reproducibly without requiring evaluation-time human annotation or LLM-based judgment. Rules are equally weighted within each aspect, and threshold-based checks may miss improvements that do not cross a predefined threshold or may penalize contextually appropriate code. The resulting NFI scores reflect performance under the specified rule set rather than a complete assessment of software quality. The rule set can be extended to incorporate additional measurable NFI criteria.

**Internal Validity.** The stochastic nature of coding agents may introduce variability into the experimental results. To mitigate this threat, we set the temperature to zero whenever supported and repeat each agent-task configuration five times. We report average results across the repeated runs to reduce the influence of run-level randomness. Manual decisions during benchmark construction may also introduce subjectivity, particularly when validating candidate tasks and constructing task descriptions. To mitigate this threat, two authors independently validate candidate tasks, and disagreements are resolved through discussion until consensus is reached. Task descriptions are drafted by one author and subsequently reviewed and refined by a second author for clarity and self-containment.

**External Validity.** Our benchmark focuses exclusively on Python projects. Although Python is widely used and has a rich open-source ecosystem, the findings may not directly generalize to other programming languages or software ecosystems. The benchmark also has an uneven task distribution across NFI aspects and task scopes. For example, it contains 90 Documentation tasks but only one multi-file Logic Patterns task. As a result, findings for sparsely represented subsets should be interpreted with caution. Future studies may extend the benchmark to additional programming languages and broaden the coverage of underrepresented task categories.

## 8 Conclusion

In this paper, we present SWE-NFI, a benchmark for evaluating coding agents on non-functional improvements beyond functional correctness. SWE-NFI contains 188 tasks constructed from real merged Python pull requests and operationalizes five NFI aspects into 92 executable rules. Its evaluation suite combines task-specific functional preservation checks with deterministic rule-based NFI evaluation. Our results show that, among outputs that pass the functional preservation checks, coding agents can achieve measurable non-functional improvements, but performance varies substantially across aspects. In particular, all evaluated agents report lower mean logic-pattern improvements than the Human Reference mean. SWE-NFI provides a reproducible basis for evaluating coding agents on non-functional software maintenance tasks beyond functional correctness and provide insights for the development of future coding agents with stronger non-functional improvement capabilities.